\documentclass[showpacs,prl,twocolumn,superscriptaddress,letter]{revtex4}
\pdfoutput=1

\usepackage{amsmath}
\usepackage{amssymb}
\usepackage{color}
\usepackage{dcolumn}
\usepackage{graphicx}
\usepackage{epstopdf}           % must come after graphicx
\usepackage{latexsym}
\usepackage{times}
\usepackage{ifpdf}
\input{epsf.sty}

\newcommand{\scri}{\ensuremath{\mathcal{J}^+}}

\newcommand{\rad}{\ensuremath{\text{rad}}}

\begin{document}
%\linenumbers
%\switchlinenumbers
%\modulolinenumbers[2]

\title{The Asymptotic Falloff of Local Waveform Measurements in Numerical
Relativity}

\author{Denis Pollney}
\affiliation{
  Max-Planck-Institut f\"ur Gravitationsphysik,
  Albert-Einstein-Institut,
  Potsdam-Golm, Germany
}

\author{Christian Reisswig}
\affiliation{
  Max-Planck-Institut f\"ur Gravitationsphysik,
  Albert-Einstein-Institut,
  Potsdam-Golm, Germany
}

\author{Nils Dorband}
\affiliation{
  Max-Planck-Institut f\"ur Gravitationsphysik,
  Albert-Einstein-Institut,
  Potsdam-Golm, Germany
}

\author{Erik Schnetter}
\affiliation{
  Center for Computation \& Technology,
  Louisiana State University,
  Baton Rouge, LA, USA
}
\affiliation{
  Department of Physics \& Astronomy,
  Louisiana State University,
  Baton Rouge, LA, USA
}

\author{Peter Diener}
\affiliation{
  Department of Physics \& Astronomy,
  Louisiana State University,
  Baton Rouge, LA, USA
}
\affiliation{
  Center for Computation \& Technology,
  Louisiana State University,
  Baton Rouge, LA, USA
}

\date{\today}

\begin{abstract}
  We examine current numerical relativity computations of
  gravitational waves, which typically determine the asymptotic
  waves at infinity by extrapolation from finite (small) radii. Using
  simulations of a black hole binary with accurate wave extraction at
  $r=1000M$, we show that extrapolations from the
  near-zone are self-consistent in approximating
  measurements at this radius, although with a somewhat reduced accuracy.
  We verify that $\psi_4$ is the
  dominant asymptotic contribution to the gravitational energy
  (as required by the peeling theorem) but
  point out that gauge effects may complicate the interpretation of
  the other Weyl components.
\end{abstract}

\pacs{
04.25.dg,  % Numerical studies of black holes and black-hole binaries
04.30.Db,  % Wave generation and sources
04.30.Tv,  % Gravitational-wave astrophysics
04.30.Nk   %Wave propagation and interactions
}

\maketitle

\noindent\textit{I. Introduction.} -- Numerical relativity has made
great strides in recent years in the solution of the binary black hole
(BH) problem. Since the original breakthroughs by
Pretorius~\cite{Pretorius:2005gq} and the moving puncture
approach~\cite{Baker:2005vv, Campanelli:2005dd}, the calculation of
long, accurate gravitational waveforms (GWs) has become an almost
routine procedure.  It is particularly satisfying that a variety of
methods (numerical methods, formulations of the Einstein equations,
wave extraction techniques) are in use and have been shown to produce
consistent results (eg.~\cite{Hannam:2009hh}).

Certain systematic errors, however, are difficult to estimate.  In
particular, current methods measure GWs at finite radii and
extrapolate the results to infinity. This extrapolation has been
identified as one of the largest remaining sources of systematic error
within current extraction techniques, particularly during the merger
and ring-down~\cite{Scheel:2008rj, Boyle:2009vi}.  Potential
ambiguities arise particularly at small radii where gauge as well as
nonlinear near-zone effects may dominate the expected polynomial
falloff of the amplitude.

In this paper, we verify the extrapolation procedure for GWs by
performing accurate wave extractions at large radii, out to $r=1000M$
from the source (where $M$ is the mass of the spacetime).  The
waveforms have been calculated using a new hybrid
multi-patch/mesh-refinement algorithm, which allows for an efficient
discretisation of the wave zone so that high accuracy can be obtained
to large radii. We find that the measured waves between $r=100M$ and
$r=1000M$ are convergent and of good enough quality to extrapolate the
phase and amplitude accurately by low-order polynomial expansions. The
measurements at $r=1000M$ can be estimated to within $0.04\%$ in
amplitude and $0.001\rad$ in phase, if the measurements out to
$r=600M$ are used in the extrapolation. This is true over the course
of the evolution, including 8 orbits of inspiral, the merger, and
ring-down. If only measurements within $r=220M$ are used, as is common,
then the errors increase by an order of magnitude.

Finally, we note that the gravitational radiation is normally
associated with the leading order term in the falloff of the
spacetime curvature. By the peeling theorem, we expect this
to be a polynomial in $1/r$ whose leading coefficient is
the Weyl component $\psi_4$. By measuring all of the Weyl
curvature components, we have been able to establish their
respective falloff rates, and verify that $\psi_4$ is indeed
the leading order coefficient with the expected $1/r$ falloff
rate. The exponents for the $\psi_1$ and $\psi_0$ components
are less clear, however, and suggest that local gauge effects
influence their computation.

\begin{figure}
   \begin{center}
    \includegraphics[width=86mm]{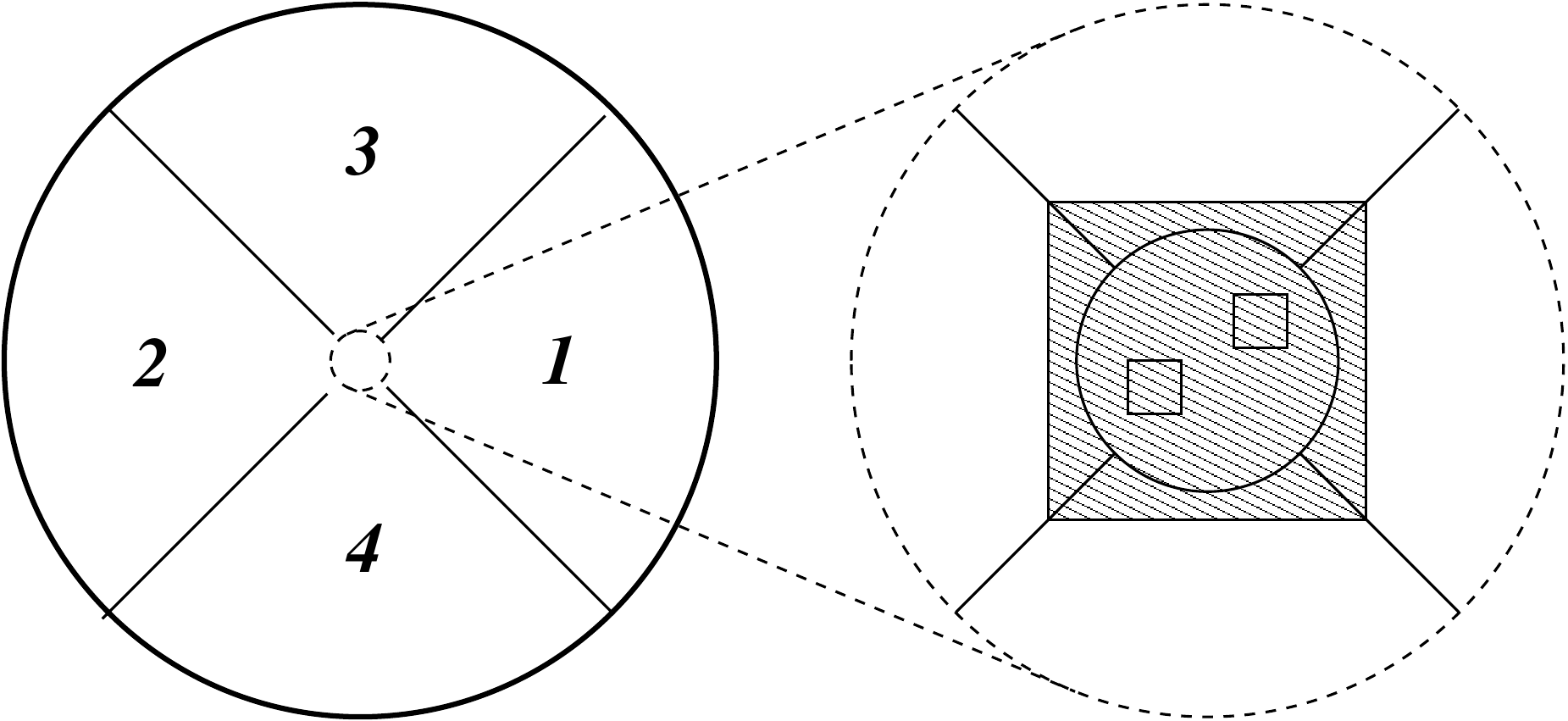}
  \end{center}
  \caption{Schematic depiction of the grid structure in the $z=0$ plane.
    Four radial grids surround the equator. The right
    shows an expanded inset of the Cartesian grid (shaded) covering the
    near-zone around the individual BHs.
  }
   \label{fig:grid}
\end{figure}

\noindent\textit{II. Computational setup.} -- A key feature of
the calculations performed here is the accuracy which we are able to
achieve at large radii from the source through the use of a newly
implemented numerical scheme. The code makes use of finite differences
and standard mesh-refinement techniques, but incorporates the use of
multiple grid patches to cover the spacetime with flexible adapted
local coordinates.  For the binary BH inspiral considered here, we
consider two regions, depicted in Fig~\ref{fig:grid}. In the near-zone
region where the BHs orbit, we discretise the spacetime using standard
Cartesian grids, applying 2:1 Berger-Oliger mesh refinement in order
to increase the resolution around each body~\cite{Schnetter-etal-03b}.
In the wave-zone, however, the dynamical fields are essentially
radially propagating waves. We cover this zone with six overlapping
patches, each of which incorporate a local radial coordinate $r$ and
transverse angular coordinates $(\rho,\sigma)$. The use of six patches
avoids the problem of a coordinate singularity on the axis of a single
spherical-polar coordinate system, as well as providing a more uniform
angular resolution over the sphere. The particular coordinates which
we have implemented are the ``inflated cube'' coordinates, given
explicitly in~\cite{Thornburg2004:multipatch-BH-excision}.

Derivatives on each grid are locally computed using standard finite
differences at 8th-order. Data is passed between patches by
interpolation, typically via centred 5th-order Lagrange
polynomials. Each patch is surrounded by a boundary zone which is
populated with data mapped from the neighbouring patch, so that
derivatives can be calculated up to the patch edge without the need
for off-centred stencils. A 4th-order Runge-Kutta integrator is used
to evolve the solution.

We write the Einstein equations in the commonly used BSSNOK form
~\cite{Nakamura87, Shibata95, Baumgarte99, Alcubierre99d}, adopting
the particular variation proposed by~\cite{Marronetti:2007wz}, whereby
the usual variable $\phi = \log\gamma/12$ is replaced by
$w=\gamma^{-2}$ (with $\gamma$ the 3-metric determinant). Gauges are
the commonly used $1+\log$ and $\tilde{\Gamma}$-driver conditions
with advection terms~\cite{Alcubierre02a, Baker:2005vv,
Campanelli:2005dd}.

The GWs are measured by evaluating the Weyl curvature tensor
components, $C_{\alpha\beta\gamma\delta}$, in a null frame
$(\mathbf{l}, \mathbf{n}, \mathbf{m}, \mathbf{\bar{m}})$, oriented so
that the outgoing vector, $\mathbf{l}$, points along the coordinate
$\hat{\mathbf{r}}$ direction while the other vectors determine an
orthonormal null frame in the local metric. The independent curvature
components $\psi_0\ldots\psi_4$ are determined by the standard
projections ~\cite{Newman62a}. The Weyl components are evaluated on
spheres of fixed coordinate radius and projected onto a
basis of spherical harmonics, ${}_{-2}Y_{lm}$. For binary black hole
situation considered here, the dominant mode is $l=2$, $m=2$, which is
used for the results presented here.

To establish the numerical accuracy, we have performed evolutions at
three different resolutions, and find that the results converge at
4th-order during the merger, and close to 8th-order during the
inspiral, in both amplitude and phase. (A more complete description of
the code implementation and tests will be given
elsewhere~\cite{Pollney:2009MP-unpublished}.) Here we present results
based on the highest resolution evolved, for which the spatial
resolution for all of the GW measurements is uniformly $h_r=0.64M$ in
the radial direction, and $h_\perp\simeq3^\circ$ in the angular
directions. GW measurements are taken every $0.144M$.

\noindent\textit{III. Extrapolation of waveforms.} -- We have evolved
an equal-mass, non-spinning binary from separation $d/M=11.0$ through
approximately 8 orbits (a physical time of around $1360M$), merger and
ring-down. The masses of the punctures are set to $m=(0.4872)$ and are
initially placed on the $x$-axis with momenta $p=(\pm0.0903, \mp
0.000728, 0)$, giving the initial slice an ADM mass $M=0.99052$. These
initial data parameters were determined using a post-Newtonian
evolution from large initial separation, following the procedure
outlined in~\cite{Husa:2007rh}, with the conservative part of the
Hamiltonian accurate to 3PN and radiation-reaction to 3.5PN, and
determine orbits with eccentricity less than $0.2\%$.

We measure the Weyl components $\psi_0\ldots\psi_4$ every $20M$ from
$r=100M$ to $300M$, then at $400M, 500M, 600M$ and $1000M$.
The radial grid
structure in the wave zone allows us to extend the outer boundary of
the grid at relatively little cost compared to Cartesian codes. For
the runs presented here, it is placed at $r=3600M$ with a resolution
of $dr=2.56M$ at the outer boundary so that the $l=2$, $m=2$ mode is
reasonably well resolved throughout the grid.  This allows for $2600M$
of evolution time before a physical or constraint violating mode
traveling at the speed of light can reach the outermost detector at
$r=1000M$. That is, the outer boundary is effectively causally
disconnected from the wave measurements presented in this paper.

The Weyl components $\psi_j = A_j e^{i\phi_j}$ are assumed to
fall-off as a function of radius according to
\begin{equation}
 \label{eq:extrapolation}
  A_j(r, t^*) = \sum_{i=0}^{n_A} \frac{A_j^{(i)}(t^*)}{r^i}, \qquad
  \phi_j(r, t^*) = \sum_{i=0}^{n_\phi} \frac{\phi_j^{(i)}(t^*)}{r^i}.
\end{equation}
The $r$ coordinate is that of the simulation coordinates, which we
find to differ by at most $0.1\%$ from the areal radius. The GWs are
expressed in terms of the retarded time $t^* = t - r^*$ where $t$ is
the coordinate time and $r^*=r + 2 M \ln [r/(2M) -1]$ is the tortoise
coordinate~\cite{Scheel:2008rj}. We do not offset the retarded time to
align the peaks of the waveforms.

It is generally difficult to estimate the error incurred when
extrapolating.  Given the data at $r=1000M$, we can attempt to gauge
an optimal choice of extrapolation parameters by attempting to
estimate this data from the measurements at smaller radii.  As test
cases, we construct extrapolations using four different sets of radii,
$e_1\in\{100M,200M\}$, $e_2\in\{160M,280M\}$, $e_3\in\{200M,300M\}$
and $e_4\in\{260M,600M\}$. Each of these incorporates 6 data points,
which over-determines low-order polynomials. We evaluate the
extrapolation coefficients by a least-squares fit to these points,
which can be important in removing spurious oscillations that may arise
fitting high-order polynomials to noisy data.

\begin{figure}
  \begin{center}
    \includegraphics[width=86mm]{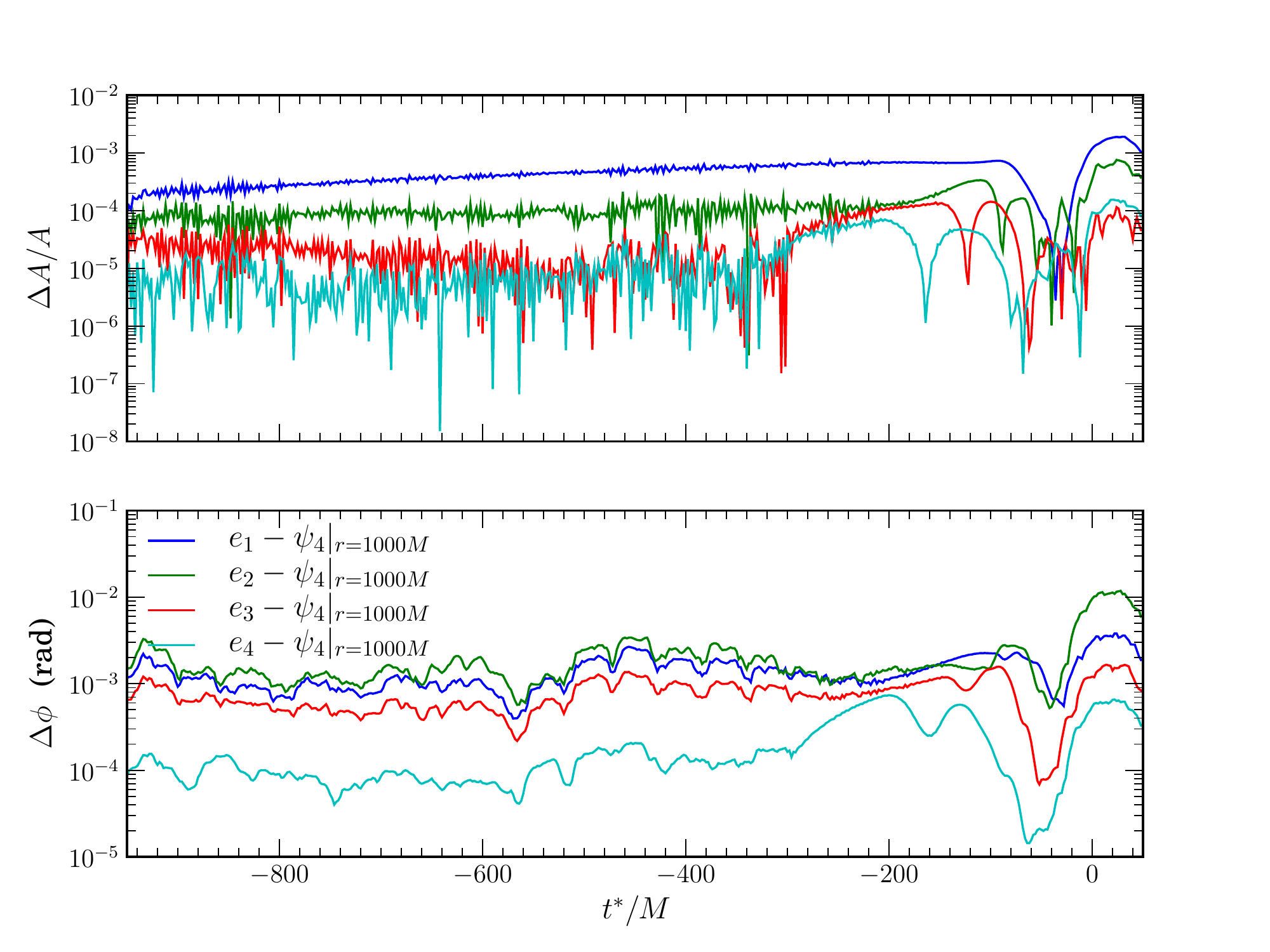}
  \end{center}
  \caption{
    Error in the extrapolated amplitude (top panel) and phase
    (bottom panel) of the $\ell=2, m=2$ component of $\psi_4$ at
    $r=1000M$ as computed by  extrapolations $e_1\ldots e_4$ (defined
    in the text).
  }
  \label{fig:extrap_1000}
\end{figure}

Fitting the amplitude using various polynomial orders, $n_A$, suggests
that in all cases $n_A=3$ is optimal in predicting the amplitude of
the measured wave at $r=1000M$, with an error of approximately
$0.02\%$ for $e_4$ and $0.2\%$ for $e_1$.  For the phase, we find that
$n_\phi=3$ minimises the error, at $6\times 10^{-4}\rad$ and $5\times
10^{-3}\rad$ for $e_4$ and $e_1$, respectively. In both amplitude and
phase, we note that the error is reduced significantly if the
outermost data $e_4$ is used. In Fig.~\ref{fig:extrap_1000} we
display the error in estimating the $r=1000M$ data using each of the
extrapolations at the optimal order. The maximum errors in both
amplitude and phase tend to occur during the late inspiral
($t^*=-200M$ to $t^*=0M$) and ring-down, although there is no sign of a
rapid growth of error during this phase.

\begin{figure}
  \includegraphics[width=86mm]{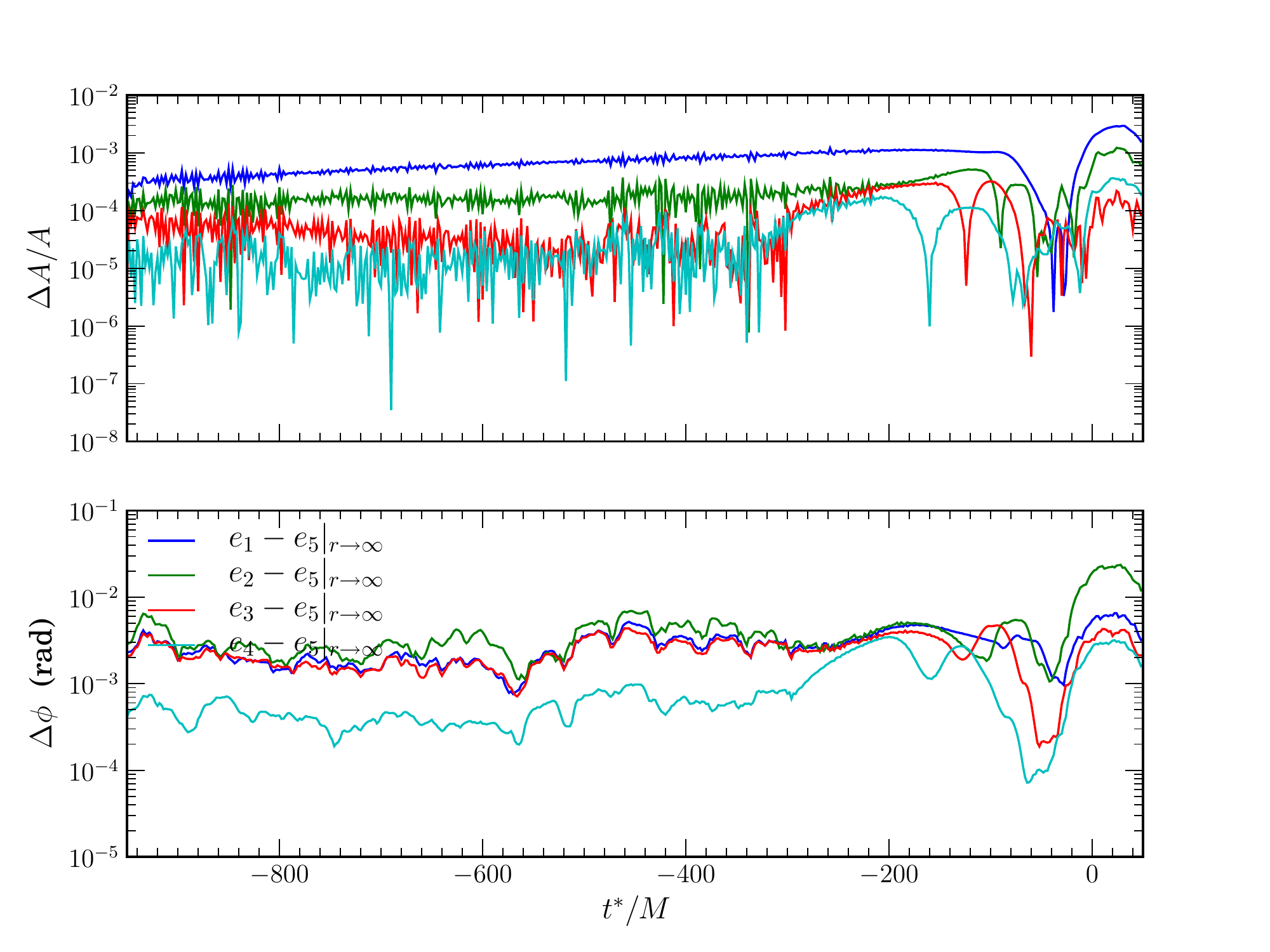}
  \caption{
    Differences between the $r\le 600M$ extrapolations
    with an extrapolation including $r=1000M$ data ($e_5$) evaluated in
    the limit $r\rightarrow\infty$.
  }
  \label{fig:extrap_infty}
\end{figure}

The corresponding extrapolations to $r\rightarrow\infty$ shows very
similar behaviour. In Fig.~\ref{fig:extrap_infty}, we have compared
each of the extrapolations with an extrapolation obtained by including
the $r=1000M$ data ($e_5\in\{280M,1000M\}$), evaluated at
$r\rightarrow\infty$. The outermost extrapolations differ by at most
$\Delta A = 0.03\%$ and $\delta\phi=0.003\rad$ over the course of the
evolution.

\noindent\textit{IV. Peeling properties.} -- The interpretation of
$\psi_4$ as the radiated gravitational energy is a consequence of
the ``peeling'' property, which states that for asymptotically
flat spacetimes at large radii, the Weyl curvature tensor can be
represented schematically as
\begin{equation}
  C_{\alpha\beta\gamma\delta} \simeq 
  \frac{\psi_4}{r} + \frac{\psi_3}{r^2} + \frac{\psi_2}{r^3}
  + \frac{\psi_1}{r^4} + \frac{\psi_0}{r^5} + O(1/r^6).
  \label{eq:peeling}
\end{equation}
That is, each component of the Weyl tensor falls off at a known fixed
rate, and at large radii, $\psi_4$ is the dominant component. At
future null infinity, $\scri$,
it can be related to $\dot{M}$, the change in energy of
the spacetime. We note, however, that the peeling theorem involves a
number of restrictions on the asymptotic form of the spacetime, and
the coordinates which are used there. A rigorous connection between
finite radius measurements and the asymptotic properties of the
spacetime at $\scri$ is difficult to make.

\begin{figure}
  \includegraphics[width=86mm]{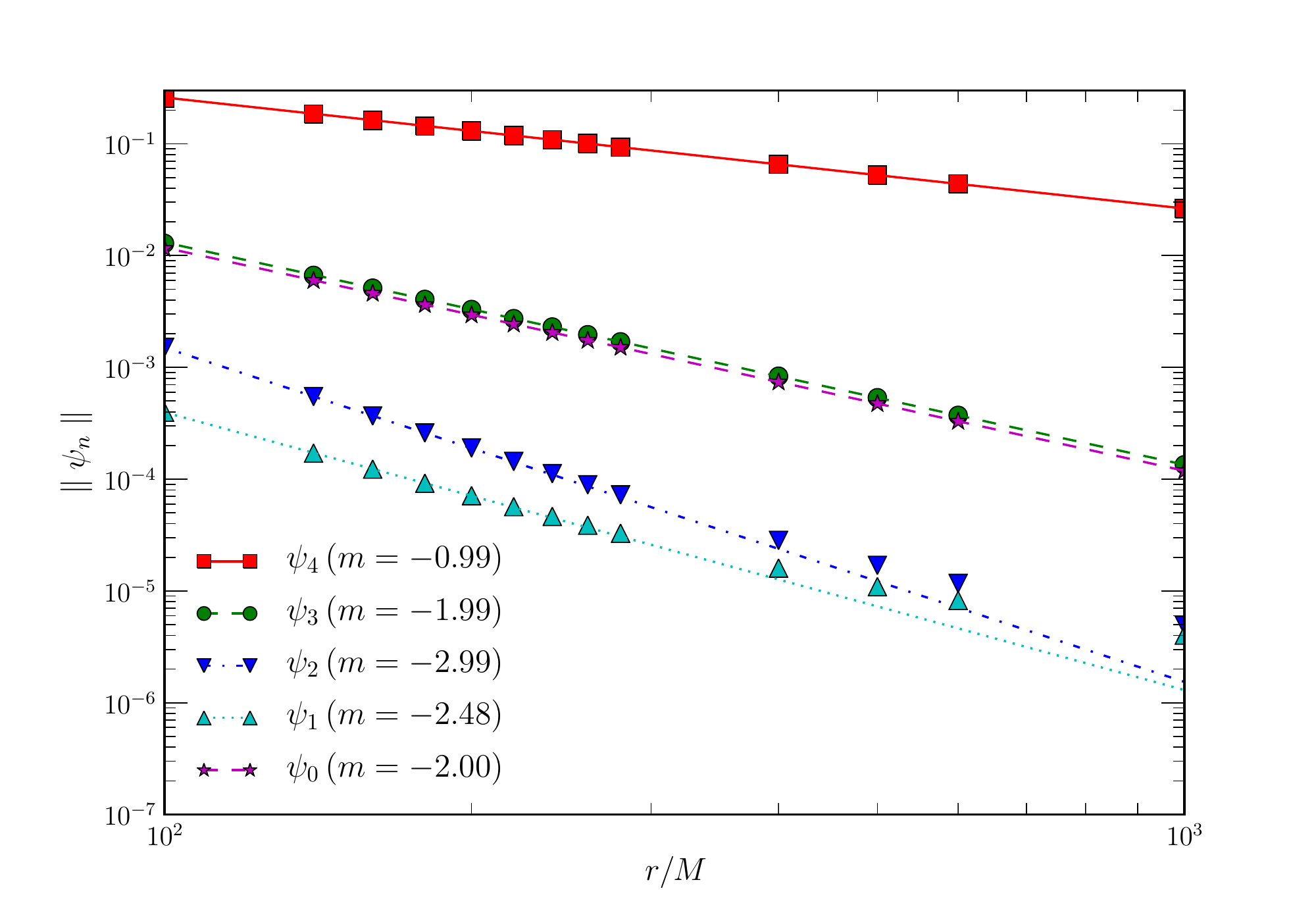}
  \caption{The radial falloff of the Weyl
    components. Lines are linear least-squares fits to all of 
    the points of $\psi_0$, $\psi_3$, and
    $\psi_4$, and the $r\le 200M$ points for $\psi_1$ and
    $\psi_2$. Measured slopes are listed in the legend.}
    \label{fig:peeling}
\end{figure}

Given the importance of the falloff of the curvature in the
identification of $\psi_4$ with the GW, it is
useful to examine the behaviour of the other Weyl components measured
by the simulation.  In Fig.~\ref{fig:peeling}, we have plotted their
falloff as a function of coordinate radius. The time series data for
each component is mapped to a scalar by integrating the amplitude over
the interval $t\in[-800M,50M]$. (Alternatively, one could obtain a
scalar by taking the measurements at a point such as the waveform
peak. A similar plot results, but the averaging effect of the integral
reduces local noise slightly.)

For the cases of $\psi_4$ and $\psi_3$, we find that a straight line
can be fitted to each of the components, indicating a consistent
exponent, with measured values of $-0.99$ and $-1.99$ respectively,
and a rather good agreement with Eq.~(\ref{eq:peeling}). Due to its
small amplitude, the $\psi_2$ measurement is dominated by numerical
noise beyond a certain radius (clear from examination of the
time-series data), and as a result, the curve veers from a straight
line. However, if we fit a straight line to the five data points from
$r\le 200M$, we find an exponent, $-2.99$, again
agreeing well with the expectation.

The $\psi_1$ and $\psi_0$ components present an interesting situation.
Particularly in the case of $\psi_0$, the amplitude is large enough
that a clear signal is present (of almost of the same amplitude as
$\psi_3$).  The falloff, however, is of order $-2.00$,
rather than the $-5$ which the peeling theorem requires.  Further, we
note that the mode propagates outwards with a peak coincident with that
of $\psi_4$, in contrast to the interpretation of $\psi_0$ as an 
``ingoing'' component of radiation.

A possible explanation is that metric perturbations cause oscillations
in the frame in which the components are measured. As described above,
we define the null frame only with reference to the local space and
time coordinates. Attempts to modify the falloff of $\psi_0$ via frame
rotations (spin-boosts and null rotations) did not preserve the falloff
of the other components.  However, other gauge effects are likely
present. We note that measurements are taken on spheres defined by the
grid coordinates. The areal radius of these spheres exhibits small (on
the order of $0.1\%$) oscillations in the $\ell=2$, $m=2$ mode. The
finite-radius $\psi_4$ measurement is known to be susceptible to pure
gauge effects such as the presence of a non-zero shift vector, which
can produce spurious GW signals in static
spacetimes~\cite{Reisswig:2009CCE-unpublished}.  Though these effects
are small, so are the values of $\psi_1$ and $\psi_0$ and thus
correspondingly sensitive compared to that of the dominant component.

\noindent\textit{V. Discussion.} -- We have demonstrated a
number of features related to the measurement of GWs
at finite radius. Our results suggest that polynomial extrapolation of
the $\psi_4$ component from small radii can provide an accurate model
for estimating the measurements at larger radii. Data measured within
$r=200M$ of the source have an error in amplitude and phase of $\Delta
A\simeq 0.2\%$ and $\Delta\phi\simeq 5\times 10^{-3}\rad$ throughout
the evolution (including merger and ring-down) compared to the
measurement at $r=1000M$. This provides an important check on
numerical relativity measurements, which typically extrapolate from
$r<200M$. Larger radius measurements do, however, improve the
extrapolation, and errors can be reduced by a further order of
magnitude if data to $r=600M$ is included. We also note that while
$\psi_4$ is dominated by the $1/r$ term beyond $r=300M$, at smaller
radii the higher order terms have a much larger contribution.

While the falloff of $\psi_4$ is the leading order contribution to the
curvature as expected, the results of Fig.~\ref{fig:peeling} suggest
that the picture may be more complicated for the other Weyl
components, and that care should be taken in interpreting local
variables according to their expected asymptotic properties. With
respect to the peeling property, we note that the asymptotic evolution
of generic initial data sets are likely to include polyhomogenous
terms, and even the situation for Bowen-York data with linear momentum
is not clear~\cite{Chrusciel:1993hx, ValienteKroon:2007ju}. While it
seems most likely that the effect we have observed in $\psi_0$ and
$\psi_1$ is related to gauge, this provides a strong caution regarding
physical predictions based on these quantities. Alternative gauge
conditions may alleviate (or exaggerate) the issues we have noted.

The mapping of finite radius results to asymptotic values at $\scri$
needs to be considered with some care.  While our results suggest that
the procedure of extrapolation is self-consistent and can be used to
estimate the results that would be obtained by direct measurement at
large radii, they do not establish the identification of the
extrapolated quantities with quantities that would be measured at $\scri$.
In a related paper~\cite{Reisswig:2009us}, we have
demonstrated that the extrapolation procedure does in fact reproduce
results obtained at $\scri$ to high accuracy, though a small
systematic error does remain.  A number of local corrections have been
proposed to improve the rigor of $\psi_4$ measurements at finite
radius (cf.\ \cite{Nerozzi:2007ai, Lehner:2007ip, Deadman:2009ds}).

Finally, we note that our accurate measurements at $r=1000M$ are a
result of a new computational infrastructure making
use of adapted coordinate grids in conjunction with a
finite difference, moving-puncture scheme. This is the first
demonstration that such methods can produce stable evolutions for
dynamical spacetimes. The efficiency gains allow the wave zone to be
covered with sufficient resolution to very large radii ($3600M$ in
this case), which has been a crucial in reducing boundary errors.

\begin{acknowledgments}
  \noindent\textit{Acknowledgments.}$-$ The authors are pleased to thank:
      Ian Hinder,
      Sascha Husa,
      Badri Krishnan,
      Philipp Moesta,
      Luciano Rezzolla,
      Juan Valiente-Kroon
    for their helpful input;
    the developers of Cactus \cite{Goodale02a} and Carpet
    \cite{Schnetter-etal-03b} for providing an open and optimised
    computational infrastructure on which we have based our code; 
    support from the DFG SFB/Transregio~7, the VESF, and by the NSF
    awards no.~0701566 \emph{XiRel} and no.~0721915 \emph{Alpaca}. 
    Computations were performed at the AEI, at LSU,
    on LONI (numrel03), on the TeraGrid (TG-MCA02N014), and the
    Leibniz Rechenzentrum M\"unchen (h0152).
\end{acknowledgments}

\bibliographystyle{apsrev-nourl}
\bibliography{aeireferences}

\end{document}